\documentclass{PoS}
\usepackage[numbers]{natbib}
\usepackage{psfig}
\usepackage{graphicx}
\usepackage{url}

\def\rg{{$R_{g}$}}

\title{The largest Swift AGN monitoring campaign: UV/optical variability in NGC 5548}

\ShortTitle{The largest Swift AGN monitoring campaign: UV/optical variability in NGC 5548}

\author{ \speaker{S. D. Connolly}$^{a,}$, I. M. M\parbox[b][2.6mm][t]{2.0mm}{c}Hardy$^{a}$, D. T. Cameron$^{a}$, T. Dwelly$^{b,a}$,
 P. Lira$^{c}$,  D. Emmanoulopoulos$^{a}$, J. Gelbord$^{d}$, E. Breedt$^{e}$, P. Arevalo$^{f,g}$, and P. Uttley$^{h}$ \\
\llap{$ˆa$} Department of Physics and Astronomy, The University of Southampton, Southampton SO17 1BJ\\
\llap{$ˆb$} Max-Planck-Institut f{\"u}r extraterrestrische Physik, Giessenbachstrasse 1, D-85748, Garching, Germany \\
\llap{$ˆc$} Departmento de Astronomia, Universidad de Chile, Camino del Observatorio 1515, Santiago, Chile\\
\llap{$ˆd$} Spectral Sciences Inc, 4 Fourth Avenue, Burlington, MA 01803 USA\\
\llap{$ˆe$} Department of Physics, University of Warwick, Coventry CV4 7AL\\
\llap{$ˆf$} Instituto de Astrof\'isica, Facultad de F\'isica, Pontificia Universidad Cat\'olica de Chile, 306, Santiago 22, Chile\\
\llap{$ˆg$} Instituto de F\'isica y Astronom\'ia, Facultad de Ciencias, Universidad de Valpara\'iso, Gran Breta\~na, 1111, Playa Ancha, Valpara\'iso, Chile\\
\llap{$ˆh$} Astronomical Institute `Anton Pannekoek', University of Amsterdam, Science Park 904, NL-1098 XH Amsterdam, the Netherlands \\
E-mail: \email{imh@astro.soton.ac.uk}, \email{sdc1g08@soton.ac.uk}
}
        
\abstract{
We report on  the largest {\it Swift} AGN monitoring program, concerning UV/optical variability in Seyferts. 
From 554 observations, over a 750d period, of the Seyfert galaxy NGC 5548, we see (McHardy et al. 2014) a good
overall correlation between the X-ray and UV/optical bands,
particularly on short timescales (tens of days).
The UVOT bands are found to lag behind X-rays with a lag scaling as wavelength to the power 1.23 +/- 0.31, in excellent agreement 
with that expected (1.33) if UV/optical variability arises from reprocessing of X-rays by the accretion disc.
However, the observed lags are $\sim$ 3 times longer than expected from a standard Shakura-Sunyaev disc, raising 
real concerns about the detailed validity of this model. The results can be
 explained with a slightly larger mass and accretion rate, and a hotter disc, or if the disc is clumpy, thereby enhancing the emission from the outer regions.
}

\FullConference{Swift: 10 Years of Discovery,\\
		2-5 December 2014\\
		La Sapienza University, Rome, Italy }

\begin{document}
\section{Introduction}
\label{sec:intro}

The relationship between UV/optical variability and X-ray variability in AGN, and the origin of the UV/optical variability, is still not clear. 
Strong X-ray/UV or X-ray/optical correlations, with short lags of less than a day, have been observed on short timescales (weeks - months)
by a number of studies \cite[e.g.][]{uttley03_5548, arevalo09, breedt09, lira11, cameron12, cameron14, shappee14}. 
However, these studies show poorer correlations on longer timescales (months - years), 
usually due to long-term  UV/optical trends not present in the X-ray variability, suggesting that 
UV/optical variability is dominated by different processes on different timescales. 
Reprocessing of X-rays into the UV/optical emission by the nearby accretion disc could cause an 
X-ray/UV correlation on short timescales, leading to UV/optical variations lagging the X-rays by the short light
travel time between the X-ray source and the disc. 
Alternatively, if X-ray variability is produced by variability 
in the UV seed photon flux due to accretion variations in the disc at very small
radii, a correlation could arise in which X-ray variability lags the UV-optical, by that same 
light travel time. Determining the precise lag between variations in the X-ray
and UV-optical emission is therefore extremely useful in determining the origin of 
UV-optical variability.

\begin{figure*}[ht!]
\centering
\vspace*{-4mm}
\includegraphics[width=0.5\columnwidth,height=0.9\columnwidth,angle=270]{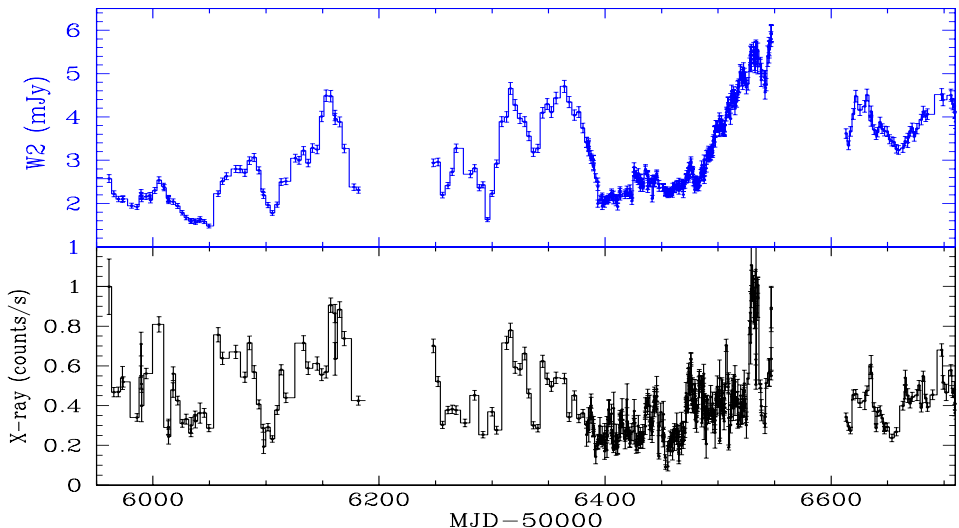}
\vspace*{-2mm}
\caption{{\it (Bottom panel)} Long term Swift 0.5-10 keV X-ray count
  rate. {\it (Top Panel)}  UVW2 flux. }
\label{fig:lclong}
\end{figure*}
\vspace*{-2mm}

Short ($\sim1$~d) lags of the X-rays by the optical are seen in
all previous studies, which mostly consist of a combination of RXTE X-ray monitoring
and ground-based optical monitoring. The lag has, however,
never been measured well enough to completely rule out an optical lead unambiguously. 
Whereas ground-based monitoring is affected by weather, the {\it Swift}
observatory can provide uninterrupted simultaneous X-ray and UV/optical
monitoring, allowing more precise measurement of wavelength-dependent lags.

Using {\it Swift} observations, the B-band was shown to lag the X-rays by $<45$mins in NGC 4395 \cite{cameron12} and
 interband lags were measured in NGC 2617 \cite{shappee14} which agree well with predictions from a reprocessing model. 

Here, we discuss the largest {\it Swift} AGN monitoring campaign to date, of 
the Seyfert 1 galaxy NGC 5548, consisting of 554 observations over a 750d period \citep{mchardy14}. 
The observations were not scheduled to follow particular events, but are instead
typical of the long-term behaviour of NGC 5548. A strong X-ray/V-band correlation was already known
to exist in NGC 5548 \cite{uttley03_5548}, but until this work the V-band lag had not yet been
precisely defined.

\begin{figure*}
\centering
\includegraphics[width=0.6\columnwidth,angle=270]{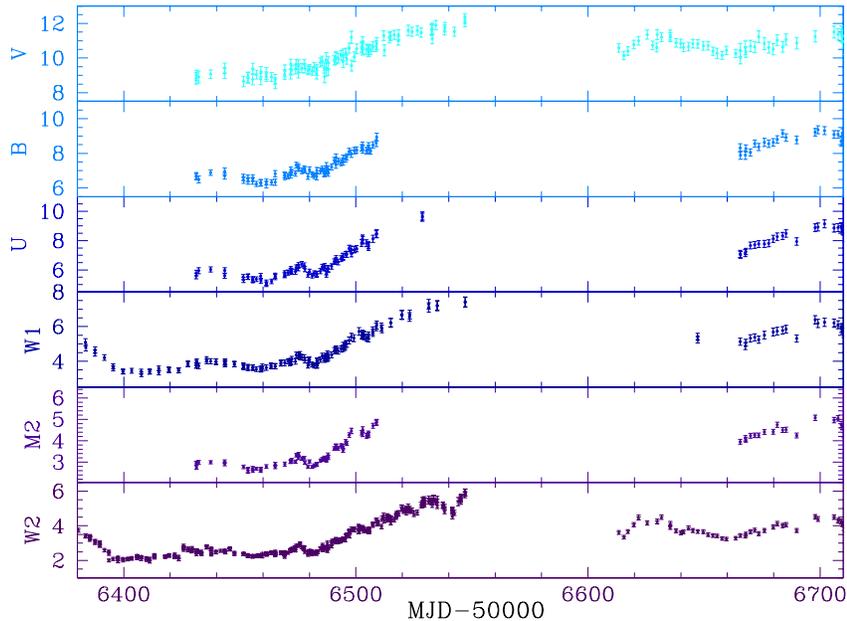}
\vspace*{-5mm}
\caption{Multiband UVOT light curves in mJy.}

\label{fig:lc_6band}
\vspace*{-4mm}
\end{figure*}

\vspace*{-4mm}

\section{Observations \& Data Reduction}
\vspace*{-2mm}

The {\it SWIFT} X-ray data were taken by the X-ray Telescope
(XRT, \cite{burrows05}) and the UV/optical observations taken by
the UV and Optical Telescope (UVOT, \cite{roming05}). All XRT
observations were carried out in photon-counting (PC) mode and all UVOT
observations were carried out in image mode. The data were analysed
using our automatic pipeline, based upon the standard Swift analysis
tasks, as described in e.g \cite{cameron12,connolly14}.  The X-ray data are corrected for
any effects caused by vignetting and aperture losses and bad pixels.  
Drops in UVOT data points were investigated individually, and those which were due
to bad tracking or bad pixels were removed.
 
Observations took place between MJD-50000 of 5960-6709, typically every 2 days, with some periods of
less or more frequent sampling (1d or 4d), and were mostly of 1~ks though sometimes of 2~ks. 
Observations were usually split into 2 or more individual visits,
improving time sampling. A total of 554 visits were made, giving, after
rejection of bad data, 465 usable X-ray measurements. 

As found previously between the X-ray and V bands \cite{uttley03_5548},
we find fairly close correspondence between the X-ray and UVW2 flux lightcurves.  On short timescales ($\sim10$d), the correspondence is strong,
but on longer timescales the amplitudes of variability are not always equal.
Observations were made in additional UVOT filters from Day 6380; 
the resultant light curves are shown in Fig.~\ref{fig:lc_6band}. 
A very close correspondence is apparent between all UVOT bands.

\vspace*{-2mm}

\section{X-ray / UV-Optical Correlations}
\vspace*{-2mm}
\begin{figure}
\centering
\includegraphics[width=0.45\columnwidth,angle=0]{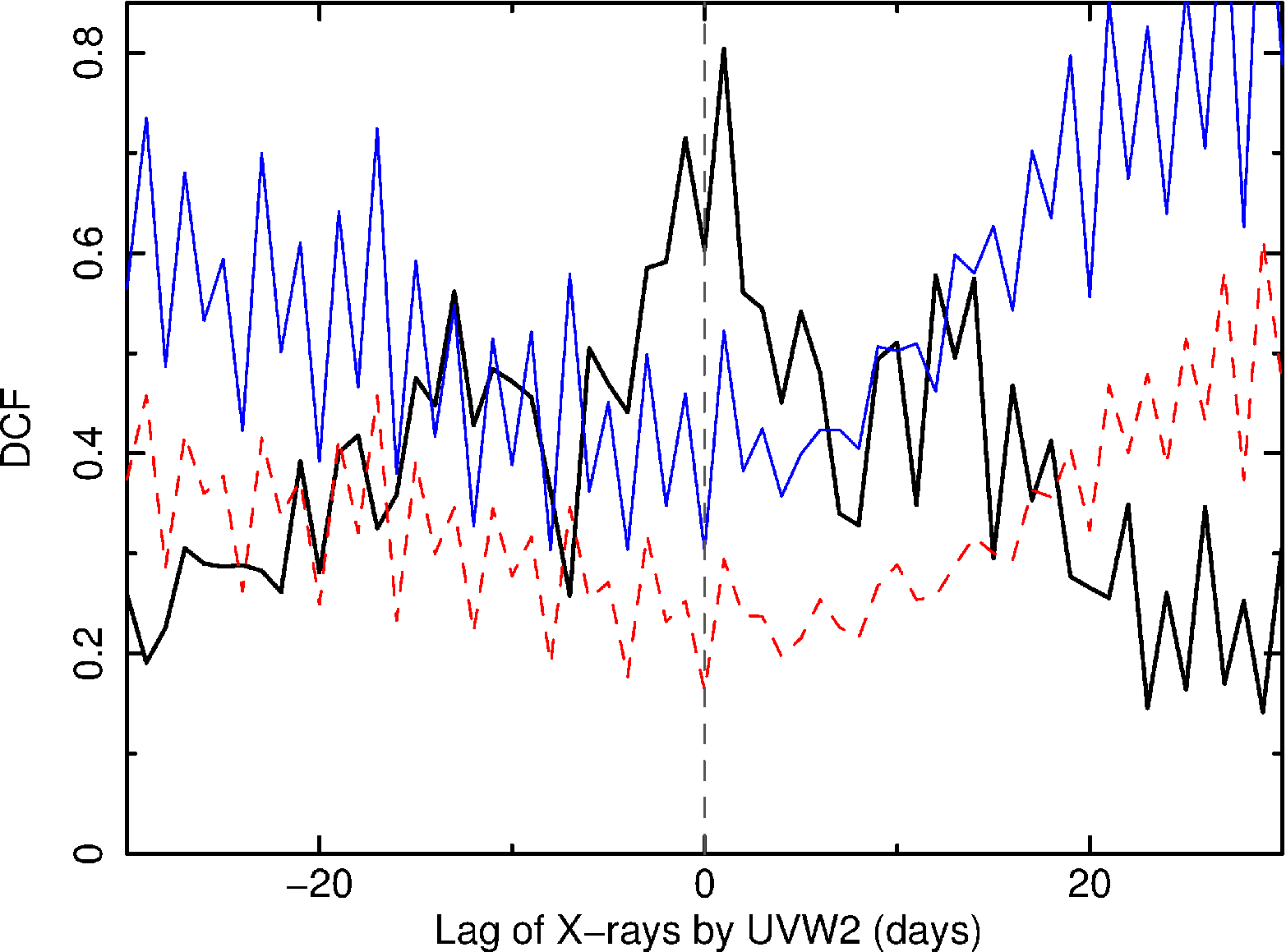}
\includegraphics[width=0.45\columnwidth,angle=0]{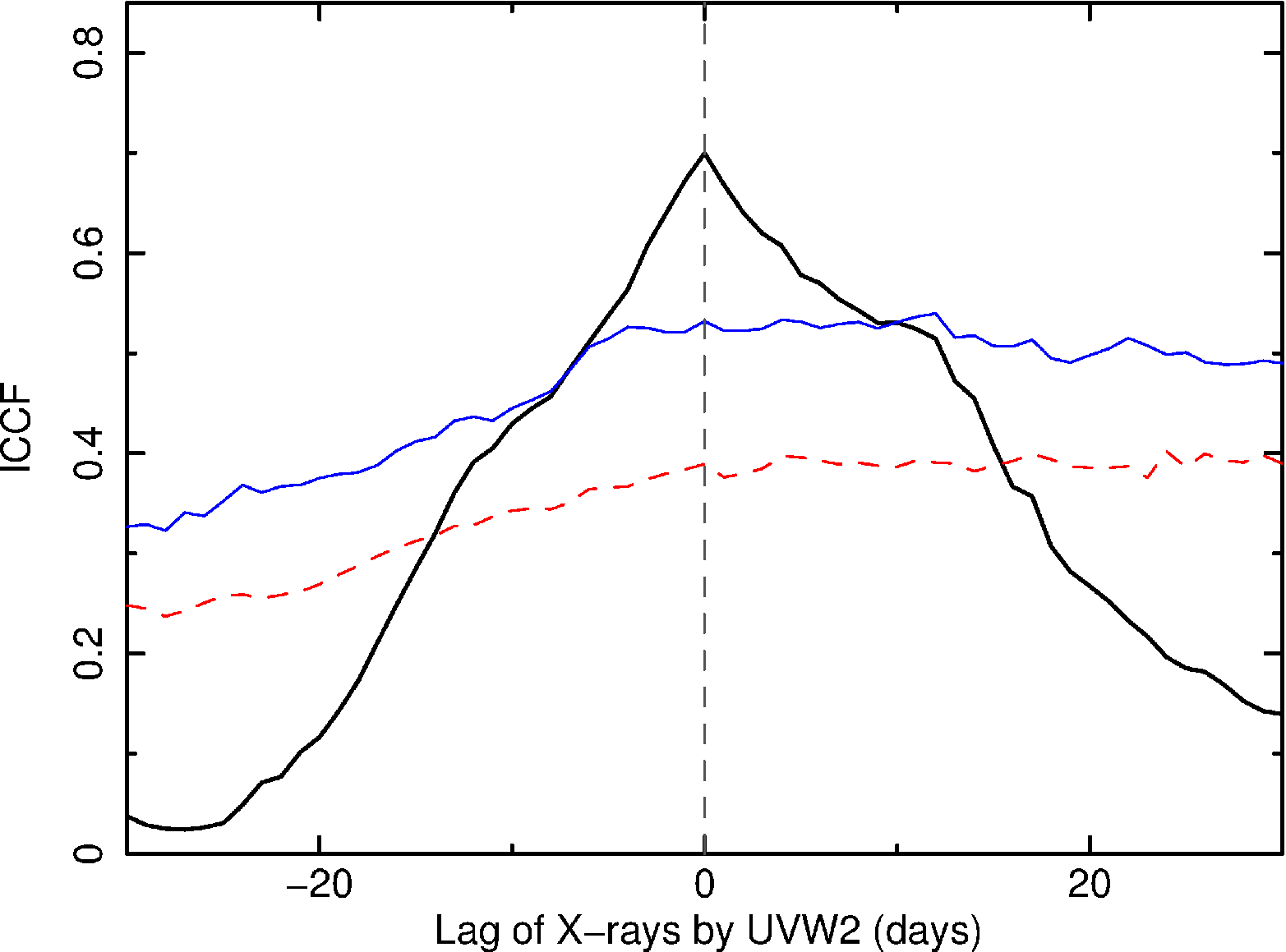}
\vspace*{-2mm}
\caption{ Left: Discrete cross correlation function between the X-ray
  and UVW2 lightcurves shown in
  Fig.~\protect\ref{fig:lclong}. The 95\% (dashed red) and
  99.99\% (solid thin blue) confidence levels are also shown.
  Right: Interpolation cross correlation function between the X-ray
  and UVW2 lightcurves shown in Fig.~\protect\ref{fig:lclong}. The 95\%
  (dashed red) and 99\% (dashed blue)
confidence levels are also shown.}
\label{fig:xw2_dcf}
\end{figure}

\subsection{The X-ray / UVW2 lag}

In Fig.~\ref{fig:xw2_dcf} we show the discrete \cite[DCF,][]{edelson88} and interpolation \cite[ICCF,][]{gaskell86} CCFs
of the UVW2 band, which is the best sampled of the UVOT bands, for the complete X-ray and UVW2 datasets.  
Only the UVW2 band, being more slowly varying, is interpolated.  
The weighted mean ICCF of the 3 sections seen in Fig.~\ref{fig:lclong} is taken, rather than interpolating across the gaps. 
The N\% confidence curves are also shown, and are defined such that (100-N)\% of correlations performed between the UVW2 data and
randomly simulated X-ray lightcurves with the same variability properties as the X-ray data would exceed these levels \citep[e.g.][]{summons07_phd}, 
 (e.g. see \cite{breedt09} for more details).
Both the DCF and ICCF show a broad, but highly significant correlation, with a peak near a lag of zero, with the DCF favouring a lag of the 
X-rays by the UVW2 of about a day. 

As correlation functions can be distorted by a long term variations in the mean
level not present in both lightcurves, it is recommended practice to subtract a running mean \citep{welsh99}.  
A mean based on a running boxcar of width 20d is therefore subtracted from both UVW2 and X-ray light curves.

\begin{figure}
\centering
\includegraphics[height=0.35\columnwidth,angle=0]{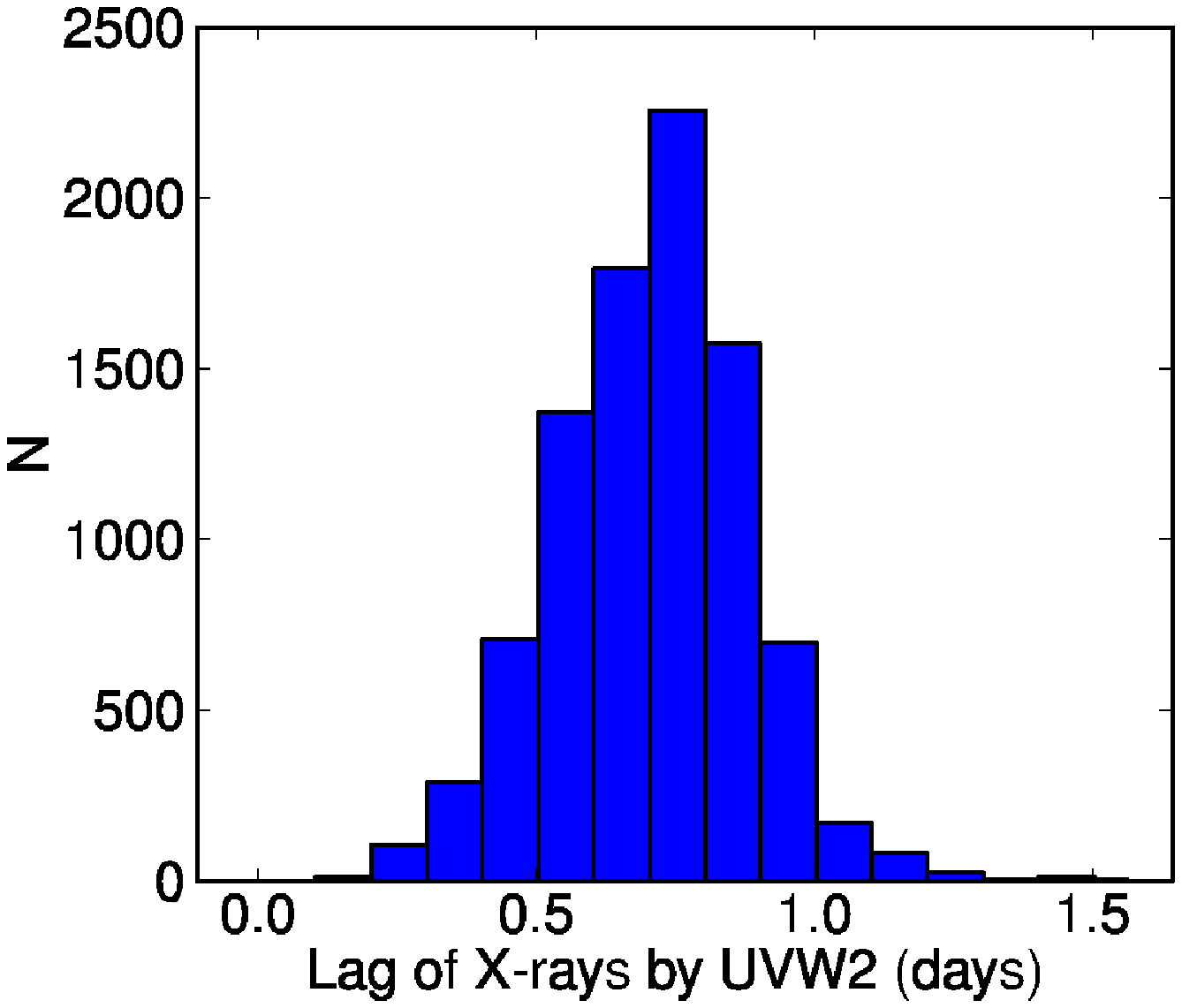}
\includegraphics[height=0.35\columnwidth,angle=0]{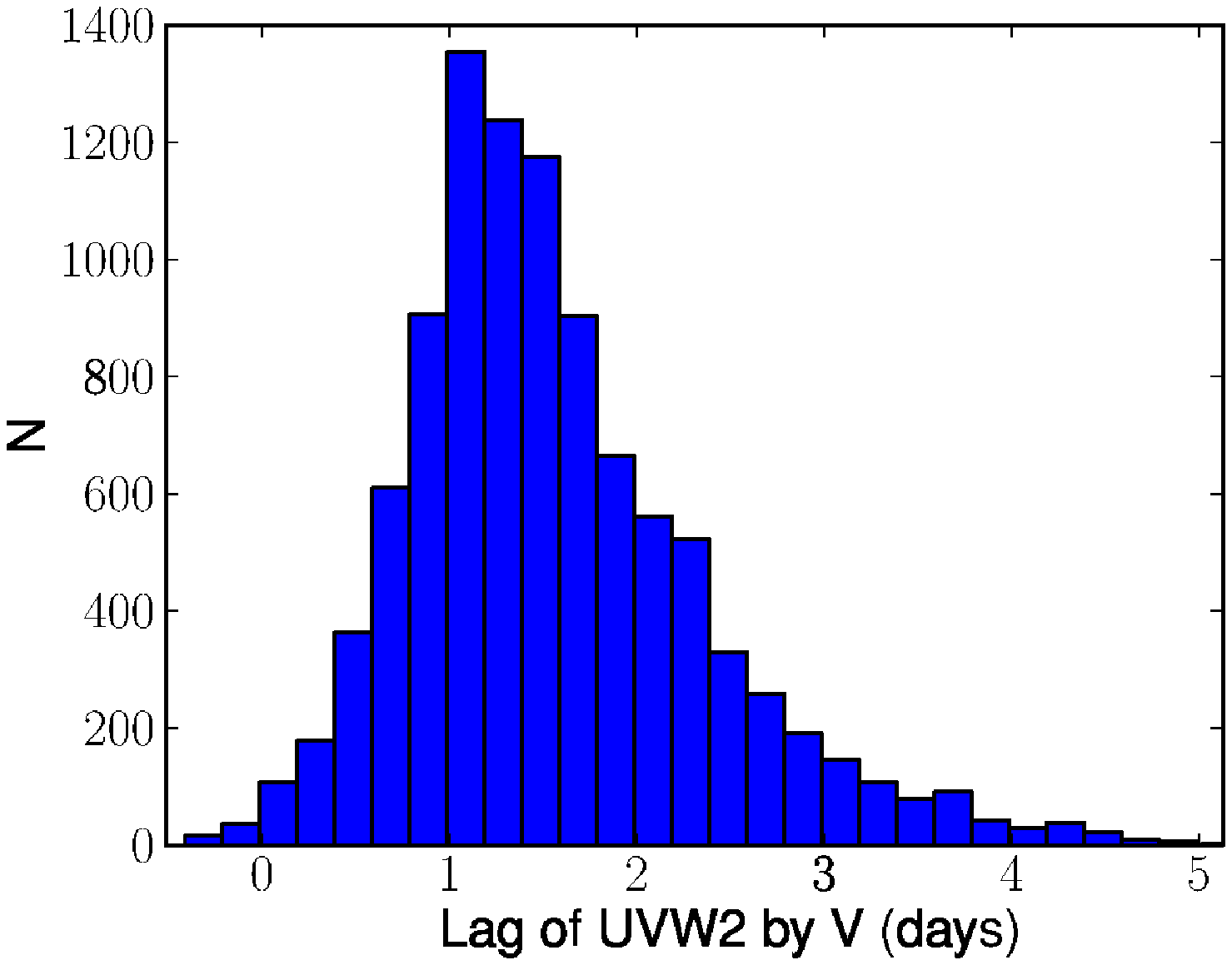}

\caption{Lag distributions from Javelin: {\it Left Panel} UVW2 following
  X-rays using data, following mean subtraction, from the intensive period from 6383-6547. 
  {\it Right Panel} V following UVW2 using data shown in Fig.~\protect\ref{fig:lc_6band}.
}
\label{fig:jav}
\end{figure}

In order to refine the lag measurement, we calculated the distribution of lags from 10000 simulations
for the period of intensive observation (Day 6383 - 6547), using the JAVELIN cross-correlation program \citep{zu11_javelin},
 used previously in e.g. \cite{shappee14,pancoast14}, allowing a lag range of $\pm10$d. 
JAVELIN assumes variability based on a damped random walk in both bands \citep{zu13_javelin}; 
although the X-ray and UVW2 variability properties are slightly different, lag measurements are not
significantly affected \citep{shappee14}.  The resultant lag distribution, shown in Fig.~\ref{fig:jav} (left panel), 
has a median UVW2 lag of +0.70$^{+0.24}_ {-0.27}$d. 
When applied to the non-mean subtracted light curves, JAVELIN does not converge to the same single distribution, 
presumably due to the uncorrelated long timescale variations.

\vspace*{-1mm}
\subsection{UVOT interband lags}
\vspace*{-1mm}

As the UVOT bands all show the same long term trends, it was not necessary to mean subtract the lightcurves. 
The lag distributions between UVW2 and the other UVOT bands were therefore calculated with JAVELIN
using data shown in Fig.~\ref{fig:lc_6band}, an example of which is shown in Fig.~\ref{fig:jav}, right panel. 
As some distributions, are slightly asymmetric, use of the mode rather than the median would slightly reduce the lags in some cases,
but the differences are small.

\vspace*{-1mm}
\subsection{X-ray reprocessing in a `Standard Disc'}
\vspace*{-1mm}

\begin{figure}
\centering
\includegraphics[width=0.5\columnwidth,angle=0]{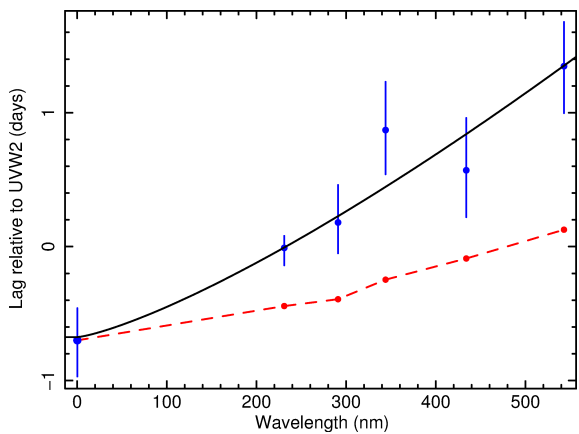}
\vspace*{-2mm}
\caption{Lag of the X-ray and other UVOT bands relative to UVW2. Lag
  $\propto$ wavelength$^{\beta}$ where $\beta = 1.23\pm 0.31$.  The
  lower, dashed, red line is the prediction for a standard disc as described
  in the text. 
 }
\label{fig:lags}
\end{figure}

On average, the UVOT bands lag the X-ray band with lags which increase with wavelength (Fig.~\ref{fig:lags}).  
These lags can be compared to predictions of reprocessing from a simple accretion disc,
for which the lag should vary as the 4/3 power of wavelength \cite[e.g.][]{cackett07,lira11}.  
The lags were therefore fitted with a simple model of the form $lag = A + (B\times \lambda)^{\beta}$.
The best fit, assuming Gaussian distributed errors, gives parameters of $A =-0.70\pm 0.21$, $\beta = 1.23\pm
0.31$ and $B=3.2 \pm 0.6 \times 10^{-3}$, with a $\chi^{2}$ of 2.1 for 3 d.o.f.  
The slope, $\beta$, is similar to that derived by \cite{shappee14} for NGC 2617 ($1.18
\pm 0.33$), but does not require an unphysical offset to the X-ray point, as they required.

This result demonstrates that the short term UV/optical
variability in NGC5548 can be very well explained by reprocessing of X-ray emission from an accretion
disc.

\section{Accretion disc Modelling}

In order to determine whether the observed lags can be explained by a disc consistent with other observed properties of NGC 5548, 
we model the disc using the methods described in \cite{lira11} (c.f. \cite{cameron14}), including the effects of both X-ray and gravitational heating.  
The key parameters in the model are the X-ray heating, obtained by extrapolating the observed 2-10 keV luminosity ($Lx_{2-10}$) to $\sim$0.01-500~keV, the disc albedo,  
the height, $H$, of the X-ray source above the disc (assuming a lamppost geometry), and the inner disc radius, $R_{in}$.

We assume a black hole mass, $M_{BH}$, of $6.7\times10^{7}$
M$_{\odot}$ \cite{bentz07}, an accretion rate, $\dot{m}_{E}$, of
$\sim 0.03-0.04$ of Eddington \cite{Vasudevan10}, a high
X-ray heating luminosity $6 \times Lx_{2-10}$ (implying a low albedo of 20\%), $H$ = 6 \rg (consistent with
X-ray source sizes measured by other methods \citep{chartas12,emmanoulopoulos14}) and $R_{in}$ = 6 \rg,
the ISCO for a Schwarzschild black hole. This produces the dashed line shown in Fig.~\ref{fig:lags}.  
The lags predicted from this standard disc model represent the time for half of the reprocessed light has been received, 
following impulse X-ray illumination. As the response is asymmetric, however, the peak response may be even faster, giving even shorter predicted model lags.

In order to increase the predicted lags from this homogeneous disc model such that they agree with observation, 
the geometry would have to be changed (e.g. $H$ = 20 \rg,
$R_{in}$ = 20 \rg), and a larger $M_{BH} = 10^{8}$ M$_{\odot}$) and
hotter ($\dot{m}_{E} = 0.06$) disc are required. In this case, disc flaring would
have a smaller effect than an increased disc temperature.  
\vspace*{-2mm}

\section{Discussion}
\vspace*{-2mm}
\subsection{Long X-ray/UV-Optical Lags} 
The form of the observed wavelength-dependent lags, i.e. lag $\mathrm{\sim \lambda^{1.23}}$ (Fig.~\ref{fig:lags}), 
strongly support reprocessing of X-rays
as the main cause of short-timescale UV/optical variability in AGN, however 
our observations also imply that this reprocessed emission originates 
further from the black hole than predicted by a `standard' accretion disc model. 
The predictions and observations can only be reconciled by pushing parameter limits. 
It is important to note, however, that a larger than expected disc is also required \cite{morgan10} to
explain microlensing observations, lending further support to the idea that the standard disc model is often inadequate,
as proposed by e.g. \cite{antonucci15}. 
The suggestion \cite{morgan10} of a low radiative efficiency is consistent with our requirement for $R_{in} \geq20$. 
Alternatively, an inhomogeneous discs, for which the outer portions contribute
more flux than for a uniform disc, would cause the disc to appear larger \cite{dexter11}.

\subsection{Long-Timescale UVOT variability} Although the correspondence of X-ray/UVW2
 is very good on short timescales, and generally good on long
timescales, from Day $\sim6470$ to 6547 there is a rise in the UV/optical
which is significantly more pronounced than in X-rays. The UV/optical rise could be
explained by an inwardly propagating rise in accretion rate, which
eventually hits the X-ray emission region, however, the viscous timescales from 100 \rg are too
long ($\sim10$~years, c.f.\cite{breedt09}). The data can therefore only be explained if the ratio
of disc scale height to radius is larger than the value of 0.1 which is 
normally assumed, or the rise propagates through a corona over the disc rather than the disc itself, 
or the perturbation starts at a small radius ($\sim 20$ \rg).

\subsection{Seed Photon Variations as a Driver of Variability}
Although the result that the X-rays always lead the UV/optical can be explained by assuming that all 
X-ray variability is generated within the corona,
it is still interesting to ask why variations in seed photon flux, which would
cause the UV-optical emission to lead, appear to have little affect
on the measured lags. We suggest that the answer is a combination of relative solid angles and conservation of photons
during the X-ray scattering process. A given single seed photon, Compton up-scattered into an X-ray photon with energies 10-100$\times$ greater than the seed photon, 
can, through reprocessing in the disc, produce many UV-optical photons by black body emission. 
As the disc fills half the solid angle seen by the X-ray source, 
a large fraction of the X-ray photons will hit the disc. 
The total number of UV-optical photons produced in this way may therefore exceed the
initial seed photon fluctuation. The reprocessing time, which is usually ignored in reprocessing calculations but which must be of finite length, 
could also add a further delay to the optical lightcurves, aiding agreement with observation.

\vspace*{-2mm}
\section*{Acknowledgements}
\vspace*{-2mm}

This work was supported by STFC grant ST/J001600/1. SDC thanks the STFC for support under a studentship. 
PL acknowledges grant Fondecyt 1120328. JG gratefully acknowledges the support from
NASA under award NNH13CH61C.


\vspace*{-2mm}

\begin{thebibliography}{99}


\bibitem[\protect\citeauthoryear{Antonucci}{2015}]{antonucci15}  Antonucci, R., 2015
\emph{Active Galactic Nuclei and Quasars: Why Still a Puzzle after 50 years?},
ArXiv, 1501.02001.

\bibitem[\protect\citeauthoryear{Ar\'{e}valo et al.}{2009}]{arevalo09} Ar\'{e}valo, P. et al., 2009
\emph{Correlation and time delays of the X-ray and optical emission of the Seyfert Galaxy NGC 3783},
MNRAS, 397, 2004.

\bibitem[\protect\citeauthoryear{Bentz et al.}{2007}]{bentz07}  Bentz, M. C. et al., 2007.
\emph{NGC 5548 in a Low-Luminosity State: Implications for the Broad-Line Region},
ApJ, 662, 205.

\bibitem[\protect\citeauthoryear{Breedt et al.}{2009}]{breedt09} Breedt, E. et al., 2009,
\emph{Long-term optical and X-ray variability of the Seyfert galaxy Markarian 79},
MNRAS, 394, 427.


\bibitem[\protect\citeauthoryear{Burrows et al.}{2005}]{burrows05} Burrows, David N., Hill, J. E., Nousek, J. A. and Kennea, J. A., 2005 
\emph{The Swift X-ray Telescope},
Sp. Sci. Rev., 120, 165.

\bibitem[\protect\citeauthoryear{Cackett et al.}{2007}]{cackett07}  Cackett, E. M., Horne, K., Winkler, H., 2007,
\emph{Testing Thermal Reprocessing in Active Galactic Nuclei Accretion Discs},
MNRAS, 380, 669.

\bibitem[\protect\citeauthoryear{Cameron et al.}{2012}]{cameron12} Cameron, D. T. et al., 2012, 
\emph{Correlated X-ray/Ultraviolet/Optical Variability in the Very Low Mass AGN NGC 4395},
MNRAS, 422, 902.

\bibitem[\protect\citeauthoryear{Cameron}{2014}]{cameron14} Cameron, D. T., 2014, 
\emph{The Relationship between UV and Optical Variability and X-ray variability in Active Galactic Nuclei},
PhD thesis, University of Southampton.


\bibitem[\protect\citeauthoryear{Chartas et al.}{2012}]{chartas12} Chartas, G. et al., 2012, 
\emph{Revealing the Structure of an Accretion Disk Through Energy-Dependent X-Ray Microlensing},
ApJ, 757, 137.

\bibitem[\protect\citeauthoryear{Connolly et al.}{2014}]{connolly14} Connolly, S. D., McHardy I. M. and Dwelly T., 2014, 
\emph{Long Term Wind-Driven X-Ray Spectral Variability of NGC 1365 with Swift},
MNRAS, 440, 3503.

\bibitem[\protect\citeauthoryear{Dexter et al.}{2011}]{dexter11} Dexter, J. and Agol, E., 2011, 
\emph{Quasar Accretion Disks are Strongly Inhomogeneous},
ApJL, 727, L24.

\bibitem[\protect\citeauthoryear{Edelson et al.}{1988}]{edelson88} Edelson, R. A. and Krolik, J. H., 1988, 
\emph{The Discrete Correlation Function: A New Method For Analyzing Unevenly Sampled Variability Data},
ApJ, 333, 646.

\bibitem[\protect\citeauthoryear{Emmanoulopoulos et al.}{2014}]{emmanoulopoulos14}  Emmanoulopoulos, D., et al., 2014,
\emph{General relativistic modelling of the negative reverberation X-ray time delays in AGN},
MNRAS, 439, 3931.

\bibitem[\protect\citeauthoryear{Gaskell \& Sparke}{1986}]{gaskell86} Gaskell, C. M. and Sparke, L. S., 1986,
\emph{Line Variations in Quasars and Seyfert Galaxies},
ApJ, 305, 175.

\bibitem[\protect\citeauthoryear{Lira et al.}{2011}]{lira11}  Lira, P. et al., 2011,
\emph{Optical and near-IR long-term monitoring of NGC 3783 and MR 2251-178: Evidence for variable near-IR emission from thin accretion discs},
MNRAS, 415, 1290.


\bibitem[\protect\citeauthoryear{Mason et al.}{2002}]{mason02} Mason, K. O. et al. , 2002,
\emph{XMM-Newton Observations of a Possible Light Echo in the Seyfert 1 Nucleus of NGC 4051},
ApJ, 580, L117.

\bibitem[\protect\citeauthoryear{McHardy et al.}{2014}]{mchardy14}  McHardy, I. M. et al., 2014,
\emph{Swift monitoring of NGC5548: X-ray reprocessing and short term UV / optical variability},
MNRAS, 444, 1469.

\bibitem[\protect\citeauthoryear{Morgan et al.}{2010}]{morgan10} Morgan, C. W., Kochanek, C. S., Morgan, N. D. and Falco, E. E., 2010, 
\emph{The Quasar Accretion Disk Size-Black Hole Mass Relation},
ApJ, 712, 1129.

\bibitem[\protect\citeauthoryear{Pancoast, Brewer \& Treu}{2014}]{pancoast14} Pancoast, A., Brewer, B. J. and Treu T., 2014, 
\emph{Modeling reverberation mapping data I : improved geometric and dynamical models and comparison with cross-correlation results},
MNRAS, 445, 3055.

\bibitem[\protect\citeauthoryear{Roming}{2005}]{roming05} Roming, P. W. et al., 2005, 
\emph{The Swift Ultra-Violet/Optical Telescope},
Sp. Sci. Rev., 120, 95.

\bibitem[\protect\citeauthoryear{Shakura \& Sunyaev}{1973}]{shakura73}  Shakura, N. I. and Sunyaev, R. A., 1973,
\emph{Black Holes in Binary Systems. Observational Appearance},
A\&A, 24, 337

\bibitem[\protect\citeauthoryear{Shappee et al.}{2013}]{shappee14} Shappee, B. J. et al., 2013,
\emph{The Man Behind the Curtain: X-rays Drive the UV through NIR Variability in the 2013 AGN Outburst in NGC 2617},
ApJ, 788, 48.

\bibitem[\protect\citeauthoryear{Summons}{2007}]{summons07_phd} Summons D. P., 2007, 
\emph{X-ray Power Spectral Densities of Active Galactic Nuclei},
PhD thesis, University of Southampton.

\bibitem[\protect\citeauthoryear{Uttley et al.}{2003}]{uttley03_5548} Uttley, P., Edelson, R., McHardy, I. M., Peterson, B. M. and Markowitz, A., 2003, 
\emph{Correlated Long-Term Optical and X-Ray Variations in NGC 5548},
ApJL, 584, L53.

\bibitem[\protect\citeauthoryear{Vasudevan et al.}{2010}]{Vasudevan10}  Vasudevan, R. V. et al., 2010,
\emph{The power output of local obscured and unobscured AGN: crossing the absorption barrier with Swift/BAT and IRAS},
MNRAS, 402, 1081.

\bibitem[\protect\citeauthoryear{Welsh}{1999}]{welsh99} Welsh, W. F., 1999, 
\emph{On the Reliability of Cross Correlation Function Lag Determinations in Active Galactic Nuclei},
PASP, 111, 1347.

\bibitem[\protect\citeauthoryear{Zu et al.}{2013}]{zu13_javelin} Zu, Y., Kochanek, C. S., Koz\l{}owski, S. and Udalski, A., 2013, 
\emph{Is Quasar Optical Variability a Damped Random Walk?},
ApJ, 765, 106.

\bibitem[\protect\citeauthoryear{Zu et al.}{2011}]{zu11_javelin} Zu, Y., Kochanek, C. S. and Peterson, B. M., 2011, 
\emph{An Alternative Approach to Measuring Reverberation Lags in Active Galactic Nuclei},
ApJ, 735, 80.

\end{thebibliography}
\end{document}